# Non-Reciprocal Response in Silicon Photonic Resonators Integrated with 2D $CuCrP_2S_6$ at Short-Wave Infrared

Ghada Dushaq[1*], Srinivasa R. Tamalampudi[1], and Mahmoud Rasras[1,2*]

[1] Department of Electrical and Computer Engineering, New York University Abu Dhabi, P.O. Box 129188, Abu Dhabi, United Arab Emirates

[2] NYU Tandon School of Engineering, New York University, New York, USA

ghd1@nyu.edu, mr5098@nyu.edu

**Abstract:**

Achieving non-reciprocal optical behavior in integrated photonics with high efficiency has long been a challenge. Here, we demonstrate a non-reciprocal magneto-optic response by integrating multilayer 2D $CuCrP_2S_6$ (CCPS) onto silicon microring resonators (MRRs). Under an applied magnetic field, the CCPS intralayer ferromagnetic ordering, characterized by easy-plane magneto-crystalline anisotropy, induces asymmetrical modal responses in the clockwise (CW) and counterclockwise (CCW) light propagation directions. The proposed configuration achieves a low insertion loss of 0.15 dB and a high isolation ratio of 28 dB at 1550 nm. Notably, it exhibits a significant resonance wavelength splitting of 0.4 nm between the counter propagation directions, supporting a 50 GHz optical bandwidth. Operating directly in the transverse electric (TE) mode, it aligns with the main polarization used in silicon photonics circuits, eliminating the need for additional polarization management. The device is ultra-compact, with a 2D flake interaction length ranging from 22 μm to 55 μm and a thickness between 39 nm and 62 nm. Its operation range covers the entire C-band with a bandwidth of up to 100 nm. These attributes make our hybrid CCPS/Si device ideal for advanced non-reciprocal optical applications in the short-wave infrared (SWIR) spectrum, crucial for enhancing the resilience of optical systems against back-reflections.

**Introduction**

Non-reciprocal photonic devices, which break the Lorentz reciprocity of light propagation, are essential for the development of advanced integrated optical devices[1–4]. These devices enable critical functionalities such as signal isolation and directional transmission, which are pivotal in applications ranging from telecommunications to quantum computing[5,6]. While electrical non-reciprocity has been effectively achieved through the use of semiconductor p-n junctions, optical non-reciprocity (ONR) presents significant challenges[7]. Achieving ONR pivots on combining innovative device design as well as the development and integration of tailored materials.

Several approaches have been explored to break Lorentz reciprocity, including nonlinear optical effects[5,8] and spatio-temporal modulation[9,10]. Among these, techniques, utilizing magneto-optical (MO) effects are particularly compelling[3,4,11–14]. Conventionally, MO devices exploit mode conversion through the Faraday effect, a common approach in bulk nonreciprocal devices[15,16]. However, on-chip waveguides, characterized by their inherent birefringence, generally favor designs that employ nonreciprocal phase shifts (NRPS)[17]. Such configurations include ring resonators, multimode interferometers, and Mach-Zehnder interferometers (MZIs)[3,4,18–22]. For waveguide integration in the short-wavelength infrared (SWIR) region, yttrium iron garnets (YIG) doped with Bi or Ce to boost Faraday rotation are particularly effective[11,23]. However, the integration of these magneto-optical garnets with semiconductor substrates faces challenges, such as substantial lattice mismatches and thermal incompatibilities[24–26]. These obstacles hinder achieving monolithic integration with strong Faraday rotation and minimal transmission loss[27]. Moreover, nonreciprocal devices utilizing these materials typically have large footprints, ranging from millimeters to centimeters, posing significant barriers to large-scale and cost-effective integration[11]. Thus, there is a pressing need for on-chip integrated nonreciprocal photonic devices that combine a suitable material with a compact footprint and robust performance.

In contrast to traditional 3D magnetic thin films, two-dimensional (2D) materials are emerging as alternatives for fabricating on-chip tunable optical components, such as electro-optic modulators, photodetectors, and filters[28–32]. Their pronounced quantum confinement effects and high refractive indices facilitate strong light-matter interactions, markedly influencing their optical properties in response to external stimuli[33,34]. Importantly, the layered architecture of 2D materials, characterized by covalent in-plane bonds and van der Waals forces between planes, enables straightforward integration onto diverse substrates, circumventing lattice matching issues and simplifying the fabrication process[29,30]. A recent study on a hybrid graphene/silicon magneto-optical isolator has been developed for integration within the SWIR region[27]. This device capitalizes on spin-orbit interactions coupled with graphene's unique properties to achieve an extinction ratio of 45 dB and an insertion loss of 12 dB at 1552 nm wavelength. However, the practical deployment of such devices is constrained by the need for cryogenic temperatures and strong magnetic fields to overcome graphene's relatively weak magneto-optical effect.

Recent advances have heightened interest in 2D multiferroic materials[35,36], notably $CuCrP_2S_6$ (CCPS), recognized for their optical transparency in the SWIR spectrum, multiple ferroic orders, and a refractive index with small contrast to silicon[37,38]. CCPS, a type-II multiferroic, intriguingly couples spin, valley, and electric dipoles. This coupling emerges from the unique interplay between its two distinct sublattices: the ferromagnetic order arises predominantly from chromium (Cr) atoms, while copper (Cu) atoms contribute to the ferroelectric polarization[39–42]. Additionally, previous studies have offered experimental and theoretical support for the presence of magnetoelectric coupling in CCPS, emphasizing the pivotal role of spin-orbit coupling in connecting electric dipoles and spins[37,40,41,43,44]. This interplay suggests that magnetic fields can induce electric polarization and that electric fields can modulate magnetic ordering, even at the monolayer level. While much research has focused on CCPS's memristive properties for data storage and neuromorphic computing applications[42,45,46], its potential in integrated photonics for efficient and chip-scale non-reciprocal devices remains underexplored. Integrating CCPS could significantly advance the development of photonic circuits by leveraging its magneto-optical properties to manage light directionally.

In this study, we explore the magneto-optic response of multi-layer CCPS integrated onto SiPh microring resonators (MRR) at the SWIR regime. Leveraging the intralayer ferromagnetic ordering within CCPS, characterized by easy-plane magnetocrystalline anisotropy, we clarify the physics underpinning significant enhancements in magneto-optical properties and non-reciprocal optical phenomena. This magnetic anisotropy enables modal asymmetry under a magnetic field, optimizing the device for non-reciprocal behavior with minimal optical losses. The CCPS-loaded MRR shows strong light-matter interaction and depicts a low insertion loss of 0.15 dB with an isolation ratio of 28 dB. The results are obtained at 1550 nm wavelength using a compact interaction length of just 22 µm to 55 µm and a 2D flake thickness ranging from 39 nm to 62 nm. The device supports a 50 GHz optical bandwidth (resonance splitting of 0.4 nm) and maintains low dispersion across a broad operational wavelength range of up to 100 nm, surpassing traditional Faraday-effect-based devices.

Furthermore, its operation in the TE-polarization mode is aligned with the primary polarization used in silicon photonics. This performance eliminates the need for additional components like polarization rotators, thus simplifying integration and enhancing efficiency. Therefore, these features position the CCPS/Si integration as a robust solution for SWIR nonreciprocal devices and optical isolators, crucial components for mitigating back-reflection in advanced optical systems.

## Results

### 1. Non-Reciprocal Photonics: Overview & Principles

In magneto-optical integrated devices, nonreciprocal phase shifts (NRPS) primarily result from field asymmetry in materials configured under Voigt geometry[17,47,48]. Figure 1a illustrates this geometry where the magnetization of the magneto-optic medium is perpendicular to the direction of light propagation. This unique alignment influences the light's behavior as it traverses the medium in different directions, primarily due to the magnetic field's impact on the material's optical properties highlighted by the material's anisotropy and mathematically expressed using the permittivity tensor[49,50].

NRPS varies with the light's direction and polarization; for instance, the behavior of the transverse magnetic (TM) and transverse electric (TE) modes depends on whether the magnetic field aligns in-plane or out-of-plane with respect to the waveguide. This leads to different propagation constants for the forward (β+) and backward (β-) traveling light[51]. The discrepancy between these constants (Δβ = β+ - β-) changes according to the direction of magnetization relative to the light propagation. This variability introduces a fundamental aspect of reconfigurability in our study. The influence of the waveguide's structural design on the NRPS modifies the propagation constants for TM and TE modes, and this dynamic is mathematically detailed as follows[11,52]:

$$\Delta\beta(\text{TM}) = \frac{2\beta^{TM}}{\omega\varepsilon_0 N} \iint \frac{\gamma}{n_0^4} H_x \partial_y H_x \, dxdy \quad (1)$$

$$\Delta\beta(\text{TE}) = \frac{2\omega\varepsilon_0}{\beta^{TE} N} \iint \gamma E_x \partial_x E_x \, dxdy \quad (2)$$

Here, the propagation constants for the fundamental transverse magnetic (TM) and transverse electric (TE) modes are denoted by $\beta^{TM}$ and $\beta^{TE}$, respectively. The angular frequency of the wave is represented by $\omega$, while $\varepsilon_0$ signifies the vacuum dielectric constant. The term $N$ refers to the power flux along the direction of light propagation, labeled as the z-direction. The parameters $n_0$ and $\gamma$ denote the refractive index and the off-diagonal component of the permittivity tensor of the magneto-optic (MO) materials, respectively. Additionally, $Hx$ and $Ex$ correspond to the magnetic and electric field components of the fundamental TM and TE modes, respectively. These parameters collectively describe the electromagnetic behavior within the MO materials under investigation.

Utilizing the Voigt configuration, we investigate the integration of a magneto-optic CCPS material within a microresonator. The CCPS is positioned to partially cover an arc of the resonator, rendering it magneto-optically active. The operational principle of this nonreciprocal optical resonator involves using the magneto-optical effect to eliminate the frequency degeneracy typically observed between the forward and backward-propagating light in optical resonators. In standard optical resonators without magneto-optical properties, the resonant frequencies for both propagation directions remain the same[53]. However, by incorporating the nonreciprocal phase shift (NRPS) effect, the resonance wavelengths become directionally dependent.

For further clarity, refer to Figure 1b. In a straight optical waveguide made of magneto-optical materials, the TE polarized modes undergo an NRPS when an external magnetic field is applied perpendicular to the light propagation direction. This shift leads to distinct effective indices ($n_{\text{eff}}$) for modes traveling forward and backward. As illustrated in Figure 1b, when this effect is applied within a ring resonator structure it results in distinct resonant wavelengths for the clockwise (CW) and counter-clockwise (CCW) directions. This effect is depicted in Figure 1c highlighting the resonance wavelength split (RWS) between CW and CCW light propagation. The RWS split not only reveals the differences in effective indices ($\Delta n_{eff}$) but also shows a divergence in their resonance wavelengths. The magnitude of the RWS ($\Delta \lambda$) is directly related to the alterations in the light propagation constant ($\Delta \beta$) and the effective refractive index, as follows[13,18,25]:

$$\Delta \beta = \frac{2\pi . \Delta \lambda}{\text{FSR} . L_{MO}} \quad (3)$$

$$\Delta \lambda = \lambda (\Delta n_{eff})/n_g \quad (4)$$

Where $L_{MO}$ refers to the interaction length of the MO waveguide, FSR is the free spectral range of the micro ring resonator and $n_g$ is the group index.

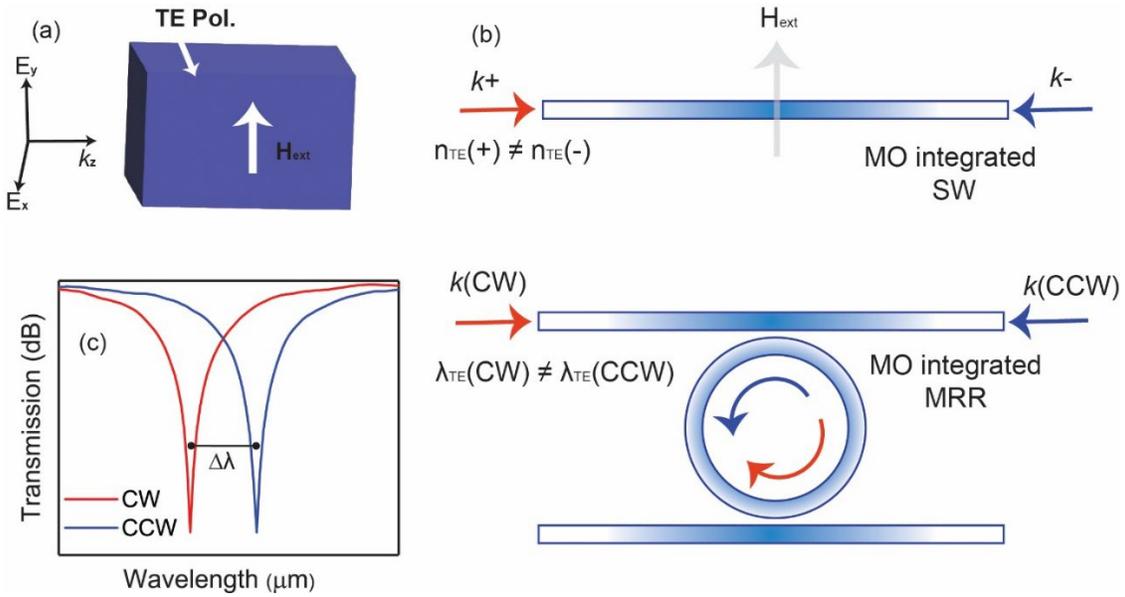

**Figure 1. Non-reciprocal phase shift (NRPS) in photonic structure (a)** Voigt configuration, illustrating the wave number ($k_z$) of light propagating in the z-direction within a magneto-optic medium **(b)** Schematic of a nonreciprocal magneto-optical waveguide depicting non-degenerate effective indices for forward and backward propagating TE polarized light, alongside a nonreciprocal magneto-optical resonator showing non-degenerate resonant frequencies between clockwise and counter-clockwise TE optical modes **(c)** Transmission spectra of the integrated magneto-optic material in a ring resonator, demonstrating the resonance wavelength split ($\Delta \lambda$) under a static magnetic field.

2. **Device design and fabrication**

Figures 2a and 2b illustrate the architecture for the hybrid integration of multilayer CuCrP$_2$S$_6$ (CCPS) with silicon photonics (SiPh) circuits. The structure utilizes silicon-on-insulator (SOI) wafers, comprising a 220 nm top silicon device layer over a 2 µm buried oxide layer. The add-drop microring resonator (MRR) features a 460 nm-wide waveguide with a 45 µm ring radius. A 100 nm gap between the bus waveguide and the resonator allows for the coupling of light, with critical coupling achieved at the through port output, optimizing the extinction ratio. In this configuration, an application of an external magnetic field perpendicular to the plane of the ring resonator induces a non-reciprocal phase shift.

The integration of CCPS onto the ring waveguide enables precise control of light propagation through the "add" and "through" ports by modulating the CCPS's refractive index via magneto-optic tuning. Light in the waveguide is evanescently coupled to the CCPS layer, and the coupling strength is modulated by the CCPS thickness.

The multilayer CCPS flakes were obtained through mechanical exfoliation from commercially available bulk crystals sourced from HQ Graphene. Comprehensive structural characterization, including AFM, EDX, XPS, and TEM with diffraction patterns, as well as optical properties such as refractive index (n), extinction coefficient (k), bandgap, and material absorption, have been detailed in our previous work[38]. Additional information is provided in the supporting information (Figs. S1-S4). For device integration, CCPS flakes with varying thicknesses ranging from ~39 nm to ~127 nm was systematically transferred onto the microring resonator (MRR) structure using a deterministic dry transfer process[29,30,54,55]. Details of the fabrication process can be found in the method section and supplementary note 4 on material integration and device testing.

Figure 2c presents the morphology of the hybrid Si/CCPS MRR via a transmission electron microscopy (TEM) cross-section image. The CCPS flakes align well with the underlying photonic structure, promoting effective light/CCPS coupling. The inset in Figure 2c reveals a thin layer of $SiO_2$ at the interface between the Si waveguide and CCPS, attributed to the oxidation of Si prior to the CCPS transfer.

To quantitatively determine the thickness of the transferred flakes, atomic force microscopy (AFM) was employed, with results shown in Figure 2d. Additional insights into AFM scans, cross-section images, and 3D reconstructed images highlighting strain effects in our devices can be found in our earlier work[38].

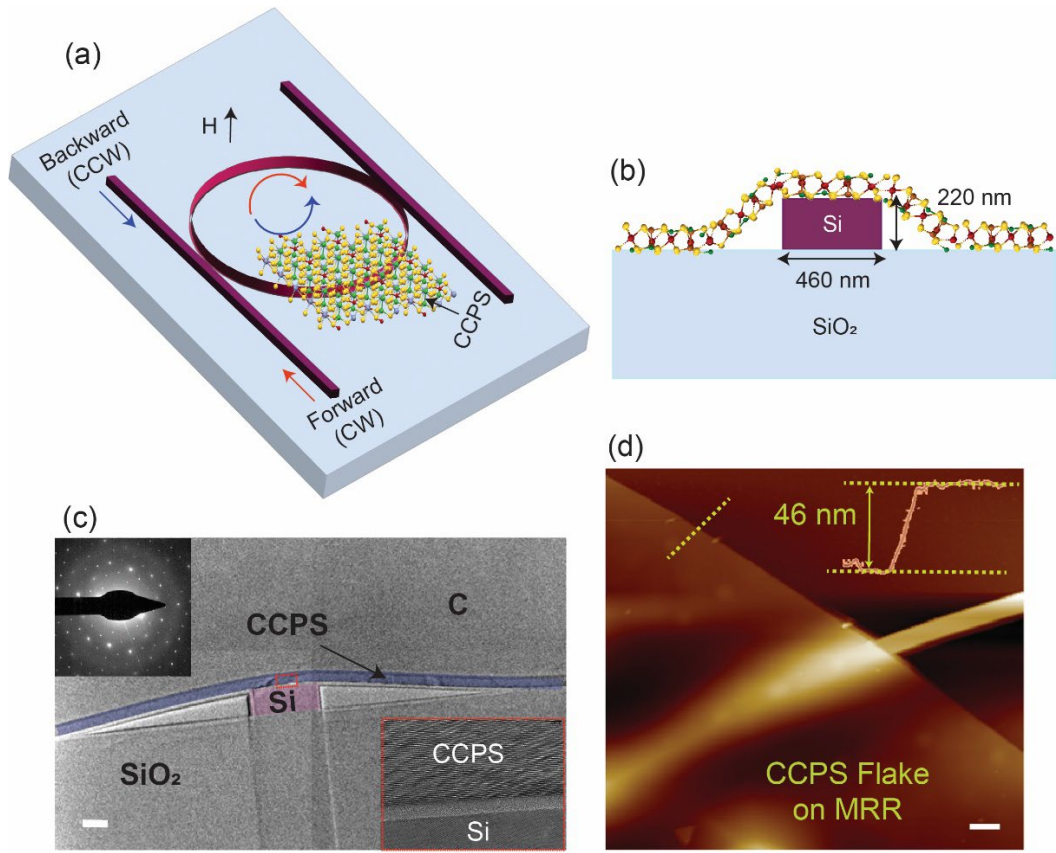

**Figure 2. Device design and CCPS/SiPh integration (a)** 3D schematic of the silicon microring resonator (MRR) design with integrated CCPS **(b)** cross-section illustrating the design and fabrication parameters; **(c)** false colored transmission electron microscopy (TEM) cross-section image of the hybrid integration with a selected area diffraction pattern (SADP) of CCPS; the red dashed square indicates the interface between the silicon waveguide and CCPS **(d)** atomic force microscopy (AFM) image scan, with the yellow dashed line showing a CCPS thickness of approximately 46 nm. Scale bars for (c) and (d) are 0.2 µm, 10 nm$^{-1}$ and 0.5 µm, respectively

## 3. Device characterization
### 3.1 Passive optical testing

To experimentally verify the non-reciprocal phase shift in hybrid integrated devices, we initially examined the passive optical transmission properties of the CCPS-loaded MRR without any magnetic field influence. These baseline measurements underpin the operational performance of the device, which we elaborate on in the subsequent section. A tunable continuous-wave laser emitting at 1550 nm shortwave infrared (SWIR) wavelengths is used to evaluate the performance of the resonator. A TE-polarized light was coupled into the MRR via a lensed fiber and the transmitted light was then collected by another lensed fiber and measured with a power meter. To control and evaluate the impact of the laser input power on the Si/CCPS chip, we systematically adjusted the input power from 0.66 mW (-1.8 dBm) to 10 mW (10 dBm). Figures 3a and 3b illustrate the transmitted resonance peak positions. As shown in these figures, the transmission spectra remain unchanged across different power levels (refer to the supporting information Fig. S5 for the transmission spectra of the bare silicon ring resonator under varying power conditions). This constancy suggests negligible thermal dissipation effects from the light propagation within the CCPS-integrated MRR. Consequently, we limited the laser power in all further measurements to a maximum of 10 mW (10 dBm). It is important to note that the 10 dBm refers to the laser input power and not the power received by the CCPS/Si-MRR, which is -2 dBm adjusted by 12 dB insertion loss per facet.

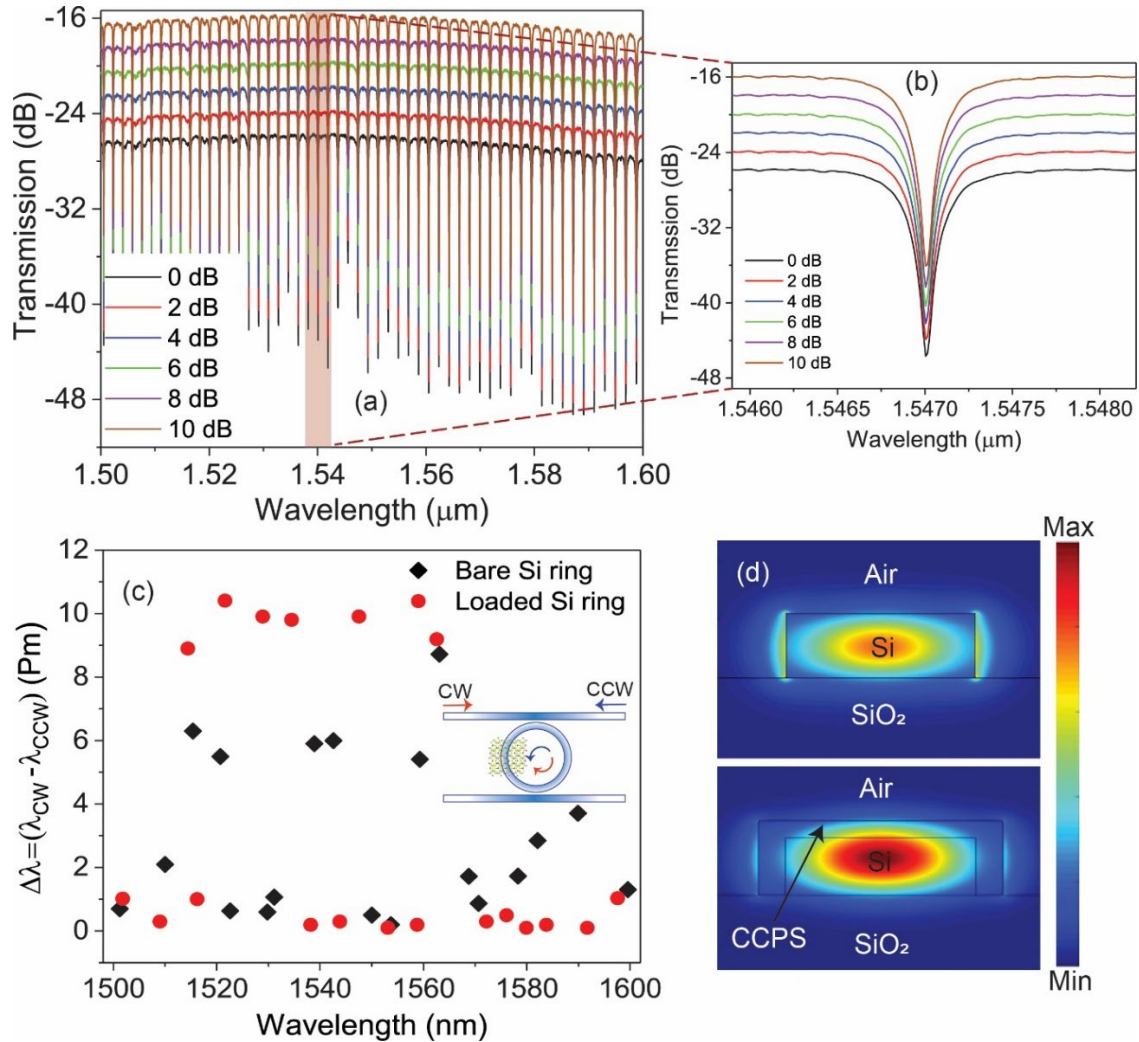

**Figure 3. Passive optical properties of hybrid Si/CCPS MRR. (a)** transmission spectra spanning 1500 nm to 1600 nm of the MRR at varying input optical powers. **(b)** detailed view of the resonance peak, illustrating minimal thermal dissipation from light propagation within the CCPS-integrated MRR. **(c)** variations in the position of the resonance peak for forward (CW) and backward (CCW) light propagation in the absence of a magnetic field. **(d)** electric field intensity profiles ($|E|^2$) for TE modes of a bare Si waveguide (top panel) and a Si waveguide with a 60 nm CCPS layer (bottom panel) at 1550 nm.

Furthermore, to establish a baseline for measurement consistency, we quantified the difference in resonance peak positions between forward (CW) and backward (CCW) light propagation in the hybrid Si/CCPS ring resonator before introducing any magnetic fields. As depicted in Figure 3c, the variation in the resonance wavelength (Δλ) between CW and CCW light propagation across a spectral range from 1500 nm to 1600 nm was assessed for both the hybrid and the bare silicon resonators. This is achieved by alternating the coupling of laser light into the chip between the input and output ports. The observed differences in both loaded and unloaded rings reached a maximum of approximately 10 ± 1 pm. Ideally, ring resonators are symmetric structures, resulting in identical transmission spectra for both clockwise (CW) and counterclockwise (CCW) propagation. However, these variations are primarily attributed to fabrication tolerances of the chip, minor temperature fluctuations, and the accuracy of peak position determination in the spectral analysis. These figures present data collected from a minimum of five devices, each featuring CCPS layers approximately 45 nm and 60 nm thick, with an interaction length of ~ 33 µm. Results from a bare silicon ring are also included for comparison.

Figure 3d illustrates the electric field intensity profiles ($|E|^2$) of the fundamental guided modes for both bare and loaded Si waveguide structures integrated with a 60 nm CCPS layer calculated at 1550 nm wavelength. The silicon waveguide predominantly supports a quasi-TE mode. Notably, the evanescent field of the mode effectively interacts with the multilayer CCPS flake, with approximately 11% of the optical mode power localized within the CCPS layer. The mode profiles for additional CCPS thicknesses are documented in Figure S6. Further details on the optical parameters measured, including calculations of the group index and effective refractive index for varying CCPS thicknesses on silicon photonics platforms, are elaborated in Fig S6.

Additional losses, induced by integrating CCPS flakes, were quantified by comparing a reference silicon ring to the hybrid CCPS/Si structure. These losses primarily stem from reflections and scattering at the interface between the passive (air/Si) and hybrid regions, intensified by mode mismatches and flake irregularities. The latter can be reduced by proper tapering between the air and the flake regions. Importantly, absorption in the shortwave infrared (SWIR) spectrum was minimal[38,39,56]. Notably, the optical losses linked to thin flakes of ~39 nm to 44 nm on the resonator were minimal, with no significant alteration in the full width at half maximum (FWHM). This balance of low optical losses and effective magneto-optic tuning surpasses the performance of other magneto-optic materials on silicon photonics platforms, typically optimized for transparency in the O- and C-optical bands for various near-infrared magneto-optic applications[18,24,25,57]. Detailed descriptions of these measurements are provided in Fig. S7-Supplement note 1 (optical loss analysis).

3.2 Magneto-optic characteristic

To investigate the magneto-optic response of CCPS, we employed a hybrid silicon-CCPS microring resonator platform. A segment of CCPS measuring 22-55 µm in arc length and 39 nm to 62 nm in thickness was integrated onto the resonator. This resonator is critically coupled to a silicon bus waveguide, rendering the transmission spectrum of the ring extremely sensitive to minor phase changes within the resonator. We applied a direct current (DC) magnetic field perpendicular to the light propagation direction across the patterned resonator using an electromagnet; detailed information on the electromagnetic setup can be found in supplementary note 2(Fig. S8). It is noteworthy that reversing the external magnetic field's direction effectively mirrors the light's propagation direction[18]. Additionally, we examined the dependency of light directionality in a bare silicon resonator under a magnetic field. As shown in Fig S9, no resonance wavelength splitting (RWS) was observed in silicon, attributed to its diamagnetic nature.

Conversely, in the CCPS-loaded ring, the effective refractive index of the optical mode is tuned by the magnetic characteristics of CCPS. Figure 4(a) presents the transmission spectra for the TE mode in a hybrid CCPS ring resonator, subjected to a magnetic field of 40 mT while being maintained at room temperature with an accuracy of ±0.2°C. This assessment was validated by repeating each measurement ten times using two methodologies: alternating the direction of light propagation and switching the direct current through the electromagnet. Notably, the transmission spectra from forward (CW-red) and backward (CCW-blue) propagation displayed alternating resonance dips, which were detuned by approximately 0.2 nm, indicating a nonreciprocal phase shift between these modes. The intralayer ferromagnetic ordering in CCPS, characterized by an easy-plane magnetocrystalline anisotropy within the van der Waals basal plane, significantly alters the magnetooptical properties. The magnetic anisotropy in CCPS is a prime factor for exhibiting non-reciprocal optical phenomena due to modal asymmetry when exposed to a magnetic field[40]. Additionally, current research has provided both theoretical and experimental evidence for magnetoelectric coupling in CCPS, highlighting the critical function of spin-orbit coupling in coupling electric dipoles and spins[40,41,43]. This interaction implies that magnetic fields can cause electric polarization and that electric fields can alter magnetic ordering. This results in a differential effective refractive index for forward and backward propagating light, thus enabling a nonreciprocal transmission spectrum essential for optical isolator functionality. Additional discussions on the magnetic properties of CCPS, including magnetic order under various temperature and magnetic field conditions, are presented in supplementary note 3 and Fig.S10.

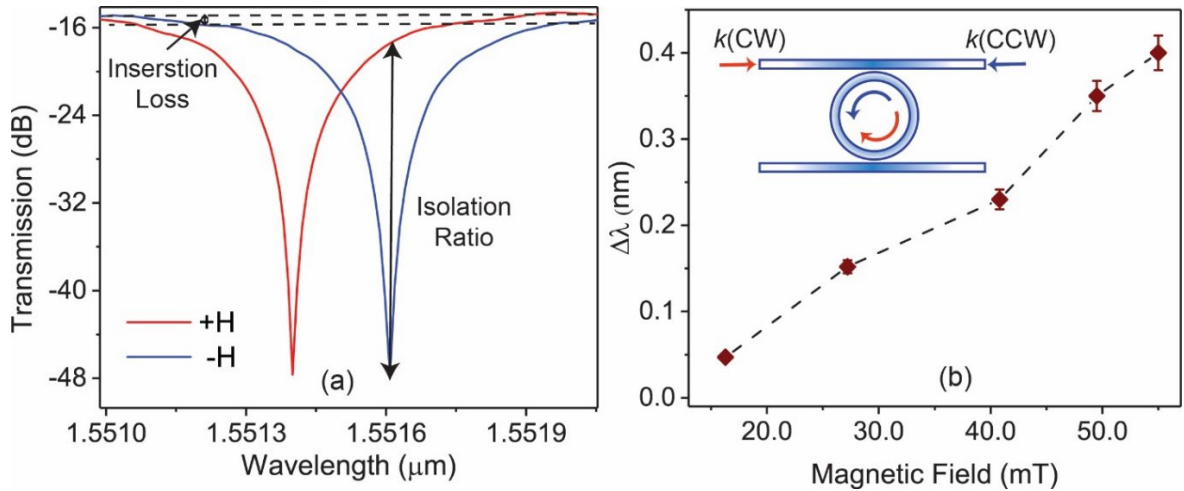

**Figure 4. Active detuning under a direct current (DC) magnetic field. (a)** transmission spectra for the TE mode, recorded with a fixed light propagation direction while reversing the direction of the magnetic field, demonstrating a non-reciprocal resonance shift. **(b)** resonance wavelength splitting (RWS), defined as $\Delta\lambda = \lambda_{CW} - \lambda_{CCW}$, plotted against the applied magnetic field strength for a device with a thickness of ~ 45 nm and interaction length of 38 μm.

At a wavelength of 1550 nm, the device demonstrated an insertion loss as low as 0.15 dB and a high isolation ratio of 28 dB, reflecting its competitive performance among monolithic magneto-optical devices[3,14,24,25,52]. Further, the resonant peak positions, recorded in both forward and backward directions across repeated tests, showed a consistent nonreciprocal resonant peak shift of (0.21 ± 0.01) nm at 40 mT. These findings are detailed further in supplement note 4, which includes comparative analyses of at least ten tested devices and evaluations of their repeatability. The quality factor and extinction ratio of the pre-fabricated rings are important determinants of the isolation ratio, with observable effects on both the isolation bandwidth and insertion loss. Refinements in these parameters may lead to improvements in overall device performance.

We also investigated the impact of varying magnetic field strengths on the resonance detuning in the device. As illustrated in Figure 4b, the device exhibited a maximum shift of approximately 0.4 nm at an upper limit of 55 mT, constrained by our experimental setup. Given the narrowband nature of the microring resonator (MR), any resonance splitting exceeding 0.1 nm facilitates isolation equivalent to the MR's full extinction ratio[18]. This was observed in our experiments, supporting a 50 GHz optical bandwidth operation. These results conclusively demonstrate the nonreciprocal light propagation in the hybrid system for the TE mode, eliminating the need for polarization rotators as previously implemented in devices that utilize magneto-optic materials[3,58].

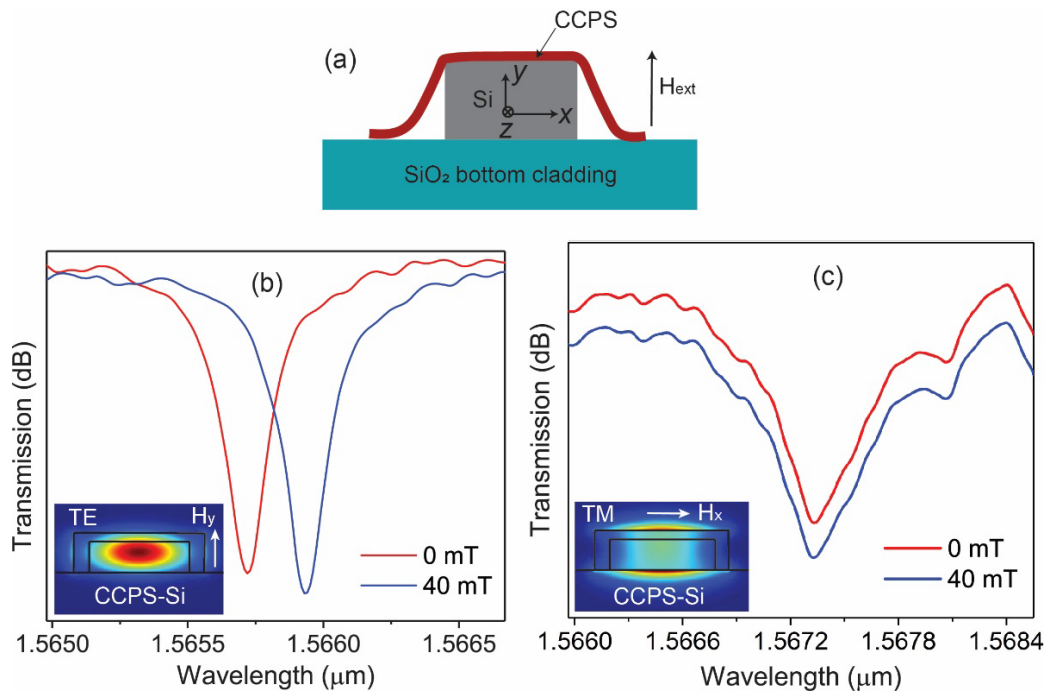

**Figure 5. Polarization-dependent characterization of the Si/CCPS hybrid device. (a)** schematic representation of the integrated CCPS/Si device with a magnetic field applied perpendicular to the direction of light propagation. **(b)** transmission spectra for the TE mode at magnetic fields of 0 mT and 40 mT, illustrating resonance detuning. **(c)** transmission spectra for TM mode at 0 mT and 40 mT, demonstrating insensitivity to the magnetic field as evidenced by the unshifted peak positions, the y-axis values have been scaled up by a factor of two to improve data visibility. Insets in (b) and (c) show the electric field distribution ($|E|^2$) in the hybrid CCPS/Si device, aligned with the direction of the magnetic field.

The observed nonreciprocal light propagation in a CCPS-integrated micro-ring resonator arises solely from magneto-optic effects, with no contribution from thermal effects. Our results, detailed in Figure S14, show that the resonance shifts are asymmetric depending on the propagating light direction. Notably, shifts induced when light is launched from the left differ in magnitude from those when launched from the right under a magnetic field, a clear deviation from the symmetric shifts expected under thermal effects[59].

Further insights are provided by the polarization-dependent measurements, illustrated in Figures 5a and 5b. The TE mode exhibits sensitivity to the magnetic field; this is attributed to the alignment of its magnetic field component ($H_y$) with the external field, as shown in Figure 5a. Conversely, the TM mode, primarily characterized by $H_x$, remains unaffected, due to its perpendicular orientation to the magnetic field. These results demonstrate that the directional and polarization-dependent behavior of the CCPS-integrated device is governed by magneto-optic interactions, highlighting the critical role of magnetic alignment in the tuning of optical properties.

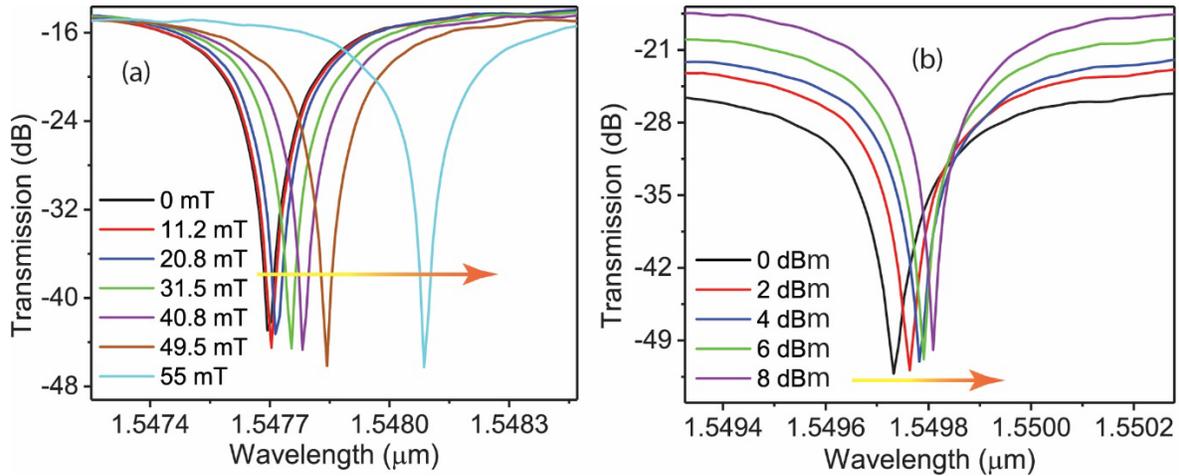

**Figure 6. Magneto-optic characteristics of the hybrid CCPS/Si micro-ring resonator (MRR).** **(a)** transmission spectra for the TE mode with a constant input laser power of 10 dBm, launched from the left, as the magnetic field varies from 0 mT to 55 mT. **(b)** transmission spectra for the TE mode under a steady magnetic field of 50 mT, with input laser power adjusted from 0 dBm to 8 dBm, launched from the left.

In our investigation of the magneto-optic characteristics of a hybrid CCPS/Si MRR, a pronounced non-linear magneto-optic response is observed which is central to the nonreciprocal light propagation. Figure 6(a) presents the transmission spectra for the TE mode with a steady input laser power of 10 dBm, launched from the left. Here, the modulation of the resonance shift across a magnetic field range from 0 mT to 55 mT is distinctly non-linear, as detailed in the supporting information (Figure S15). Additionally, Figure 6(b) shows the transmission spectra for the TE mode under a consistent magnetic field of 50 mT, with a varied input laser power from 0 to 8 dBm. Unlike the Faraday effect, which typically shows no power dependency, our results exhibit this characteristic.

These findings indicate that the Magneto-Optical Kerr Effect (MOKE) is the primary mechanism behind the nonreciprocal behavior observed in our device, rather than the Faraday effect. The measured nonlinear and power-dependent characteristics are crucial for confirming MOKE as the driving force behind the observed nonreciprocal light propagation. Additional effects such as spin-orbit coupling due to the pronounced magnetoelectric properties of CCPS can influence the behavior of our device and contribute to the observed non-linear response[60].

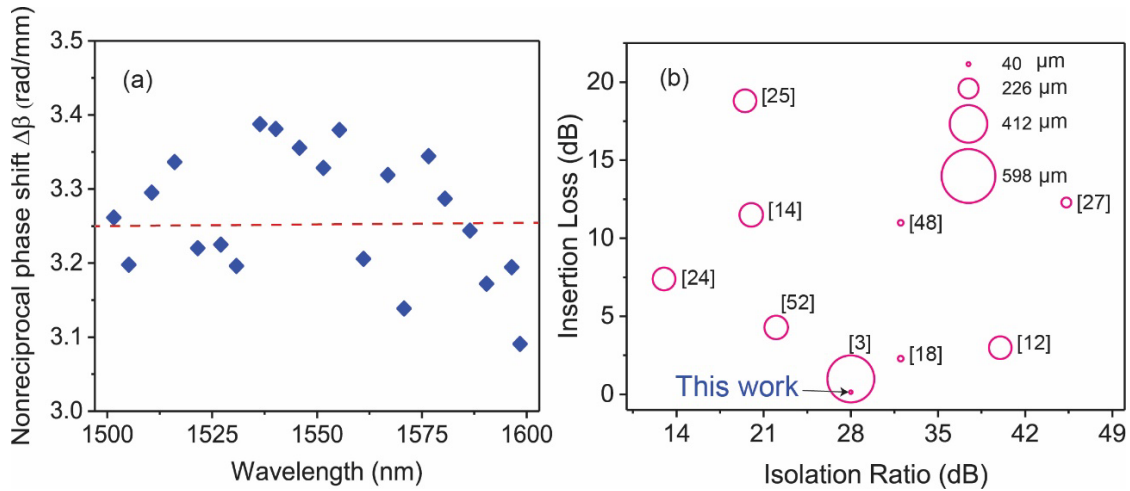

**Figure 7. Performance of the nonreciprocal hybrid CCPS/Si- MRR device**. **(a)** measured nonreciprocal phase shift (Δ$\beta$) across a range of wavelengths. **(b)** comparative analysis of research on magneto-optic (MO) materials integrated into ring resonators near the C-Band, focusing on isolation ratio, insertion loss, and MO material integration length.

Dispersion is another critical factor in determining the performance of nonreciprocal photonic devices. Therefore, we evaluated the dispersion characteristics of the nonreciprocal phase shift (Δ$\beta$) over the wavelengths from 1500 nm to 1600 nm, as depicted in Figure 7a. The dispersion exhibited by Δ$\beta$ is relatively low across the measured wavelength range. It maintained an approximate value of 3.25 rad/mm, which suggests a potential for broader operational bandwidth. This modest level of dispersion could help reduce the wavelength dependency and dispersion limitations often encountered in conventional nonreciprocal devices[13].

Furthermore, our measurements indicate a consistent nonreciprocal response extending from 1500 nm to 1600 nm, and from 1280 nm to 1360 nm, as referenced in Figure S16. The observed operational bandwidths are 100 nm and 80 nm, respectively, in these ranges. These findings underscore the enhanced performance and versatility of our hybrid CCPS/Si devices, highlighting their potential for broader applications in photonic systems.

Figure 7b displays a bubble chart comparing magneto-optic (MO) materials integrated into ring resonators near the C-Band. It outlines the isolation ratio, insertion loss, and MO material integration length across different studies. Notably, our hybrid CCPS/Si device features a compact interaction length of 22 μm to 55 μm and a thickness range of 39 nm to 62 nm, minimizing the device's footprint while maintaining low insertion losses and a high isolation ratio.

Our device operates in the TE mode consistent with the main polarization used in silicon photonics integrated circuits[61]. It is in contrast with other studies that primarily target the TM mode and often require additional components, such as polarization rotators, to achieve nonreciprocal TE mode responses[3,58]. These components typically contribute to higher insertion losses. The direct TE mode operation of our CCPS/Si device avoids these complexities, offering a simpler and potentially more efficient solution for photonic circuits, particularly in applications where future integrated lasers predominantly operate in TE modes. This functionality aligns with the practical needs of advanced optical systems.

## Discussion

The investigation into the magneto-optic response of multilayer CCPS integrated on silicon photonics reveals notable advancements in integrated non-reciprocal optical devices at near-infrared wavelengths. The intralayer ferromagnetic ordering within CCPS, characterized by an easy-plane magnetocrystalline anisotropy, significantly influences the magneto-optical properties and induces modal asymmetry under magnetic fields. This anisotropy is key for the non-reciprocal phenomena observed, making the CCPS/SiPh devices effective in balancing low optical losses with strong light-matter interaction.

The device exhibits a low insertion loss of 0.15 dB and an isolation ratio of 28 dB at a wavelength of 1550 nm. These metrics demonstrate the device's functional efficiency and competitive performance relative to monolithic alternatives. The design favors the transverse electric (TE) mode commonly used in silicon photonics circuits, avoiding the need for extra components like polarization rotators, which typically add to losses and complexity.

Additionally, the device's compact dimensions, with a 22 μm to 55 μm interaction length and 39 nm to 62 nm thickness, help reduce its footprint while maintaining functionality. The observed resonance splitting of 0.4 nm at an applied magnetic field, supporting a 50 GHz optical bandwidth, along with low dispersion allowing for operational bandwidths up to 100 nm, showcases the device's capability for broader applications.

The results demonstrate the potential of hybrid CCPS/Si devices as efficient, compact optical isolators in the short-wave infrared range, essential for reducing back-reflection in advanced optical systems. This work contributes to the ongoing development of highly integrated and functional nonreciprocal photonic devices for future optical networks.

## Materials & methods

### Materials supply

Bulk $CuCrP_2S_6$ crystals were sourced from HQ Graphene (https://www.hqgraphene.com/).

### Scanning electron microscopy (SEM) and X-ray dispersive spectroscopy (EDX)

Photonic chips were mounted on SEM stubs using carbon tape and imaged under high vacuum mode with a Quanta 450 field emission scanning electron microscope (FEI) at an electron energy of 10 kV. For compositional analysis, EDX mapping was performed on a CCPS bulk crystal also mounted on SEM stubs with carbon tape. EDX spectra were acquired at an electron energy of 30 kV.

### Atomic force microscopy (AFM)

AFM measurements were conducted using a WITec AFM module integrated with a research-grade optical microscope, operating in tapping mode. The cantilever tip used was a Scanasyst-air with a tip radius of 7 nm, a force constant of 0.2 N m$^{-1}$, and a resonance frequency of 14 kHz.

### Transmission electron microscopy (TEM)

High-resolution analytical scanning/transmission electron microscopy (S/TEM) was performed using an FEI Talos F200X operating at 200 keV. This instrument combines high-resolution S/TEM imaging with a four-quadrant energy-dispersive X-ray spectrometer (EDS) for detailed elemental and compositional mapping. The sample was prepared as a thin lamella, where e-carbon and ion carbon layers were applied as protective coatings to ensure the integrity of the lamella during imaging.

**X-ray photoelectron spectroscopy (XPS)**

The composition of the CCPS was analyzed using X-ray Photoelectron Spectroscopy (XPS) on a Nexsa G2 XPS system (Thermo Fisher Scientific). This system incorporates SnapMap™ imaging, enabling precise alignment of small X-ray spots with bond pads for accurate surface chemical analysis. Measurements were performed using a monochromatic Al Kα X-ray source.

**Magnetic force microscopy (MFM)**

Magnetic force microscopy (MFM) measurements were conducted using an NX10 atomic force microscope (AFM) system equipped with a magnetic field generator (Park Systems). The magnetic field generator applied an in-plane magnetic field with a maximum strength of approximately 335 Gauss to the sample. The MFM mode was employed with a lift height of 6 nm. The samples consisted of CCPS 2D material deposited on a $SiO_2$/Si substrate. The flakes exhibited varying sizes and thicknesses, and two relatively thin flakes were selected for detailed measurements under different applied magnetic fields.

**Mode analysis and FDTD simulation**

The electric field profile within the silicon waveguide was determined using the eigenmode solver in MODE Solutions, part of Lumerical's Device Multiphysics Simulation Suite.

**Spectroscopic imaging ellipsometer**

The optical properties of multilayer CuCrPS were assessed using an Imaging Ellipsometry system from Accurion (https://accurion.com/company). This system combines optical microscopy with ellipsometry, enabling spatially resolved measurements of layer thickness and refractive index. The technique is highly sensitive to ultrathin films, capable of analyzing structures from sub-nanometer single layers to multilayers several microns in thickness. Imaging Ellipsometry provides layer thickness measurements with a spatial resolution of up to 1 μm. The ellipsometric parameters Psi (ψ) and Delta (Δ) were evaluated and fitted using EP4 model software.

**Stamping of CCPS on the photonic chip**

Multilayered flakes were exfoliated using Scotch tape and subsequently transferred onto a PDMS substrate. Employing a precise dry transfer method, selected CCPS flakes were positioned onto the Si waveguides. Detailed information on the deterministic transfer process is provided in the supplementary material.

**Optical characterization**

The optical transmission was measured by edge coupling light into the device structure using lensed fibers connected to a tunable laser in the short-wave infrared (SWIR) band (Keysight 8164B Lightwave Measurement System). The output light from the devices was collected with an output lensed fiber and detected by a power meter. Polarization of the light (TE/TM) was calibrated using reference rings fabricated on the same chip with identical geometries. Calibration of output optical power intensities was performed with a standard photodiode power sensor prior to device testing. All optical measurements were performed in air at room temperature. The chip was mounted on an electromagnet stage, which was controlled by a DC power supply. Additional details about the experimental setup can be found in Supplementary Note 2.

### Funding

This material is based upon works supported by Tamkeen under NYUAD RRC Grant No. CG011 and the NYU Abu Dhabi Research Enhancement Fund under Grant No. REF244.

### Acknowledgment

The authors acknowledge the NYUAD Photonics and Core Technology Platform Facility (CTP) for their support in analytical, material characterization, device fabrication, and testing. We thank Dr. Qiang Zhang for assistance with characterizing the magnetic properties of CCPS. We are grateful to Dr. Paul Mack, senior application scientist at Thermo Fisher Scientific, for his help with acquiring XPS data. Additionally, we thank Dr. Rocky Nguyen from the Research Application Technology Center at Park Systems for recording the MFM data under various magnetic fields. The first author acknowledges the L'Oréal-UNESCO For Women in Science Middle East Fellowship.

### Conflict of Interests

The authors declare no competing interests.

### Contributions

G.D conceived the research idea, designed the experiments, conducted the majority of the experiments, collected and analyzed the data, and drafted the manuscript, S.T provided insights during the research process, M.R supervised the research project and critically reviewed and revised the manuscript, all authors have read and approved the manuscript.

### Additional information

Supplementary information

Supplementary information accompanying the manuscript is available at (http://www.nature.com)

Supplementary Information for

**Non-Reciprocal Response in Silicon Photonic Resonators Integrated with 2D CuCrP$_2$S$_6$ at Short-Wave Infrared**

Ghada Dushaq[1]*, Srinivasa R. Tamalampudi[1], and Mahmoud Rasras[1,2]*

[1] Department of Electrical and Computer Engineering, New York University Abu Dhabi, P.O. Box 129188, Abu Dhabi, United Arab Emirates

[2] NYU Tandon School of Engineering, New York University, New York, USA

**This file includes:**

## Structural properties of CCPS

The layered Van der Waals material CuCrP$_2$S$_6$ (CCPS) is part of the transition metal thio/selenophosphates (TPS) family, characterized by a monoclinic crystal structure (Pc space group)[1,2]. Figure S1a presents a 3D schematic (bc plane) of the material's structure. Each monolayer consists of a sulfur framework with octahedral cages occupied by Cu and Cr ions, as well as P-P pairs. Cu ions alternately occupy upper (Cu1) and lower (Cu2) positions, leading to an antiferroelectric (AFE) state at low temperatures. The crystal structure forms triangular networks composed of quasi-trigonal CuS$_3$, octahedral CrS$_6$, and P$_2$S$_6$ units. Cr ions and P-P pairs are almost centrally located within a layer, while Cu ions are slightly off-centered[1,3,4]. Figures S1b and S1c present the AFM image scan and the cross-sectional thickness measurements of a stair-like CCPS flake.

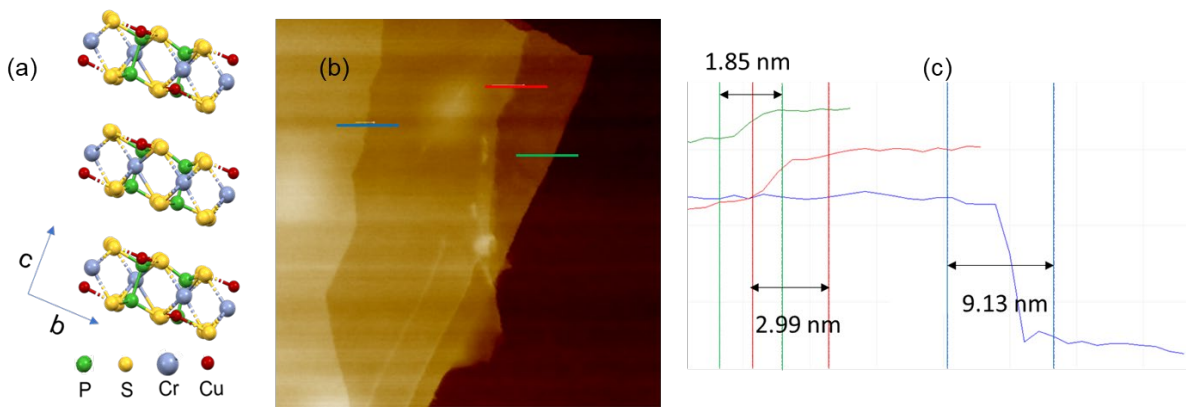

**Figure S1. Structural properties of CCPS: (a)** 3D representation (ball and stick model) of the bc plane. **(b)** AFM image scan of a stair-like CCPS flake. **(c)** Cross-sectional thickness measurements, with green, red, and blue step heights corresponding to the colored solid lines in the AFM scan.

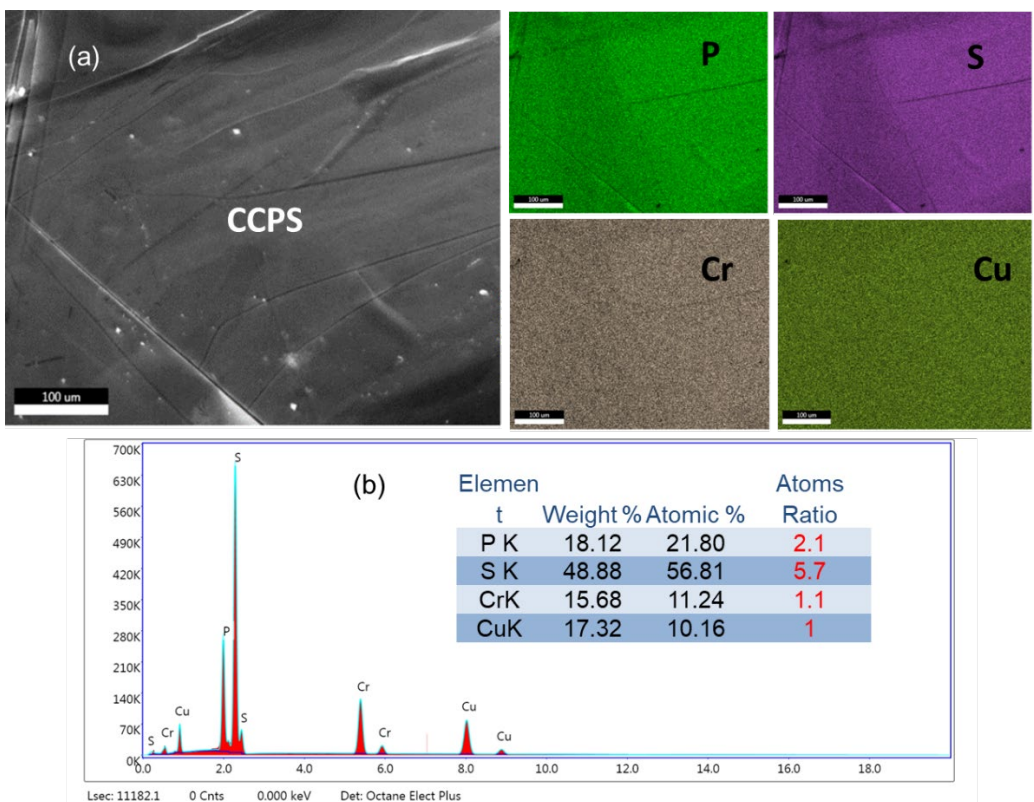

**Figure S2. Energy dispersive X-ray (EDX) analysis of a CCPS sample using Scanning Electron Microscopy (SEM): (a)** SEM image of the analyzed area with quantitative EDX mapping of CCPS. **(b)** EDX spectra displaying the detected elements, accompanied by a table presenting the atomic and weight percentage composition of $CuCrP_2S_6$. The scale bar in (a) represents 100 μm. The observed ratio of 1:1.1:2.1:5.7 closely matches the expected stoichiometric composition of CCPS (1:1:2:6).

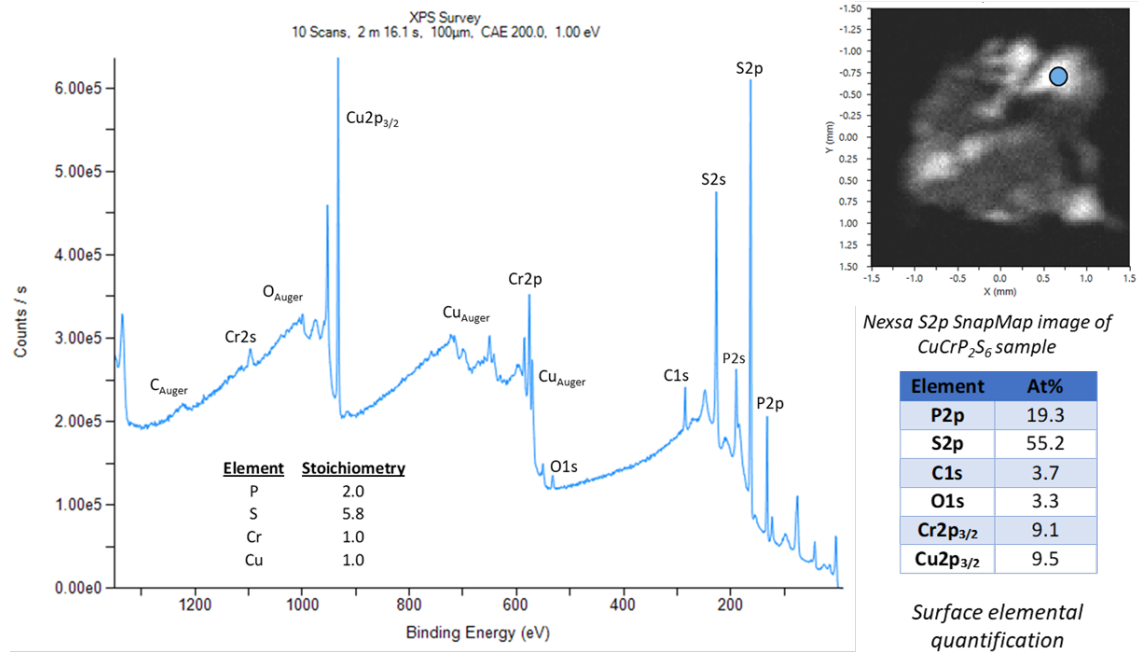

**Figure S3. X-ray photoelectron spectroscopy (XPS) data acquired using the Thermo Scientific Nexsa G2:** The spectrum is from a selected location on the SnapMap™ image (blue circle) and shows the stoichiometric composition of CCPS (1:1:2:5.8), closely matching both the EDX data and the expected stoichiometry.

.

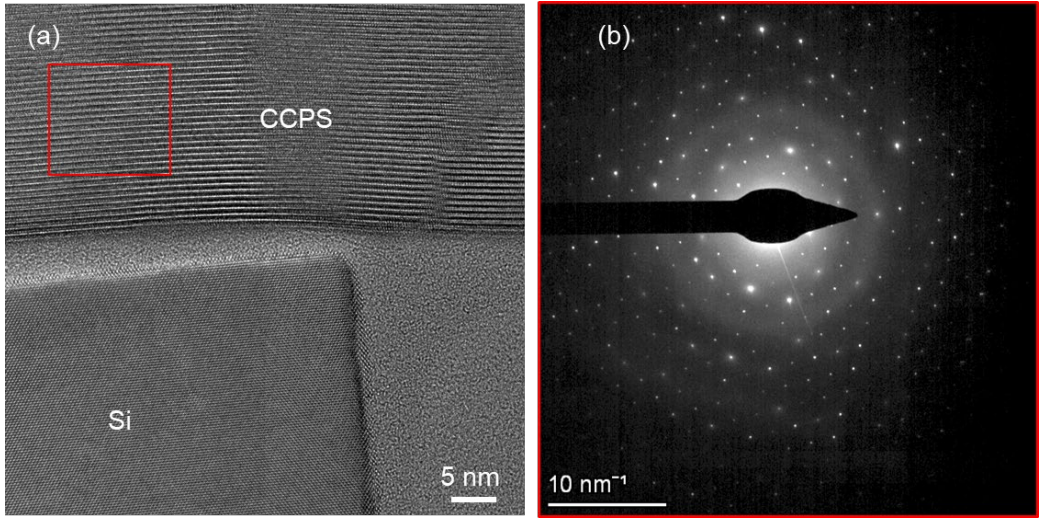

**Figure S4. Transmission electron microscopy imaging: (a)** High-resolution image of the Si waveguide and integrated CCPS interface, highlighting its crystallinity. **(b)** Selected area diffraction pattern (SADP) recorded from the red square region in (a).

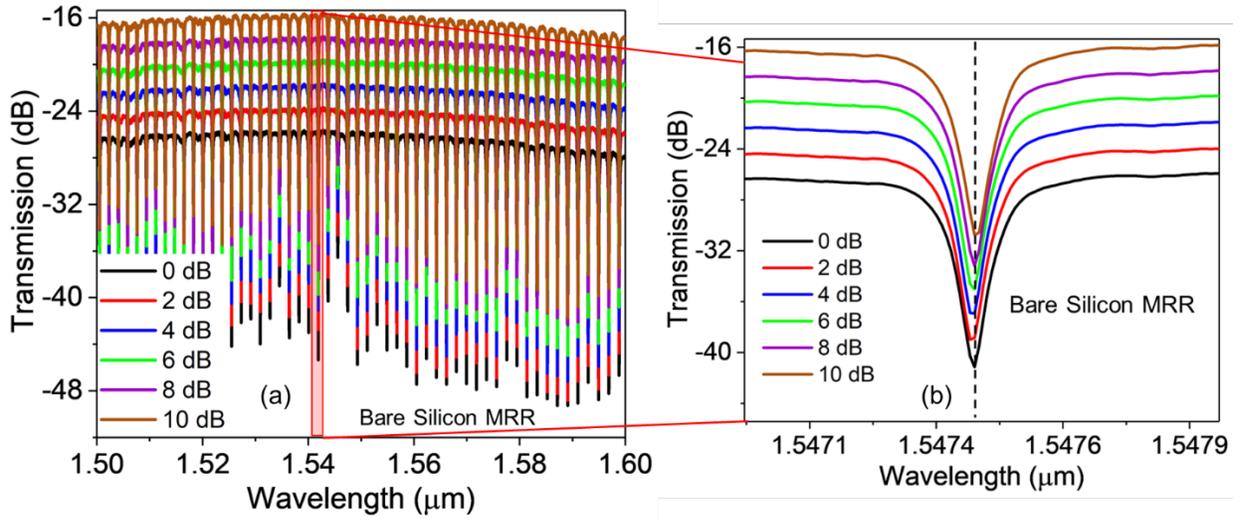

**Figure S5. Passive optical properties of bare Si MRR. (a)** transmission spectra spanning 1500 nm to 1600 nm of the Si MRR at varying input optical powers. **(b)** detailed view of the resonance peak, illustrating minimal thermal dissipation from light propagation within the Si MRR.

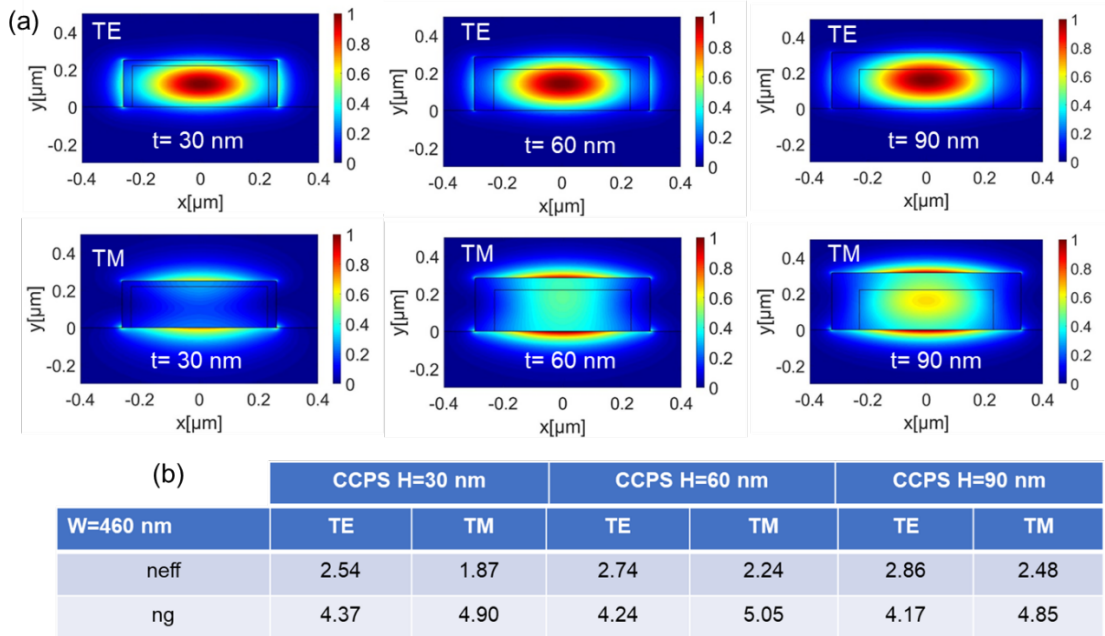

**Figure S6. Modal analysis of hybrid CCPS/Si at varying CCPS thicknesses (t): (a)** Electric field intensity profiles ($|E|^2$) for TE and TM modes of the hybrid CCPS/Si waveguide with 30 nm, 60 nm, and 90 nm CCPS layers at 1550 nm. **(b)** Extracted group and effective refractive indices for different thicknesses.

**Supplement note 1**

**Optical loss analysis in ring resonators with multilayer CCPS integration**

To quantify the change in optical loss per unit length (Δα) due to the integration of multilayer CCPS in the evanescent field region of the ring resonators, we observed the reduction in the ring's quality factor (Q)[5]. The intrinsic quality factor (Q) of the ring is related to the loss per unit length through the equation[6,7]:

$$Q = \frac{10}{\ln(10)} \frac{2\pi n_g}{\lambda_0 \alpha} \quad (1)$$

Where $\lambda_0$ is the resonance wavelength, and $ng$ is the group index.

We extracted the loaded quality factor ($\lambda_0$/FWHM) from the experimental transmission spectra using Lorentzian fitting (see Figure S7). The intrinsic quality factor ($Q_i$) can be determined from the loaded quality factor ($Q_l$) using the relationship[8]:

$$Q_i = \frac{2Q_l}{1 \pm \sqrt{T_0}} \quad (2)$$

Where $T_0$ is the normalized transmitted power at the resonance wavelength.

The device design is optimized for critical coupling conditions at 1550 nm for the TE mode, resulting in an intrinsic quality factor of approximately $2Q_l$. In the post-integration of CCPS, losses mainly arise from reflections and scattering at the coupling interface between the passive waveguide (air/Si) and the hybrid region. This is attributed to mode mismatch and flake irregularities.

The change in optical loss (Δα) can be expressed as[6]:

$$\Delta\alpha = \frac{2\pi n_g}{\lambda_0} \frac{10}{\ln(10)} \left( \frac{1}{Q_f} - \frac{1}{Q_i} \right) \quad (3)$$

Note: 10/ln (10) is the conversion from linear to logarithmic. $Q_f$ is the quality factor of the ring after CCPS integration, and $Q_i$ is the quality factor before integration. The group index $ng$ (~ 4.2) was obtained from FDTD Lumerical simulations. The quality factors before and after integration were calculated from the FWHM of the resonator's transmission spectra, as shown in Fig. S7, using Lorentzian fitting. By substituting the measured quality factor values into Eq. (3) and using Eq. (2) to obtain the intrinsic Q, the optical loss due to CCPS can be determined. This value was then normalized to the CCPS interaction length ($L_{Mo}$).

We conducted measurements on both thin (~39 nm) and thick (~127 nm) CCPS flakes. The transmission spectra for these flakes are presented in Fig. S7. For thin flakes, optical losses were found to be negligible, while for thick flakes, losses reached up to 0.018 dB/μm. This significant loss in thick flakes can be attributed to increased light scattering at the input and output regions of the CCPS, caused by the larger step height.

Our comparative analysis of multiple devices revealed that the magneto-optic effect in our case was more sensitive to the interaction length of the flakes rather than their thickness. Consequently, we optimized our devices to balance efficient magneto-optic tuning with minimal optical losses.

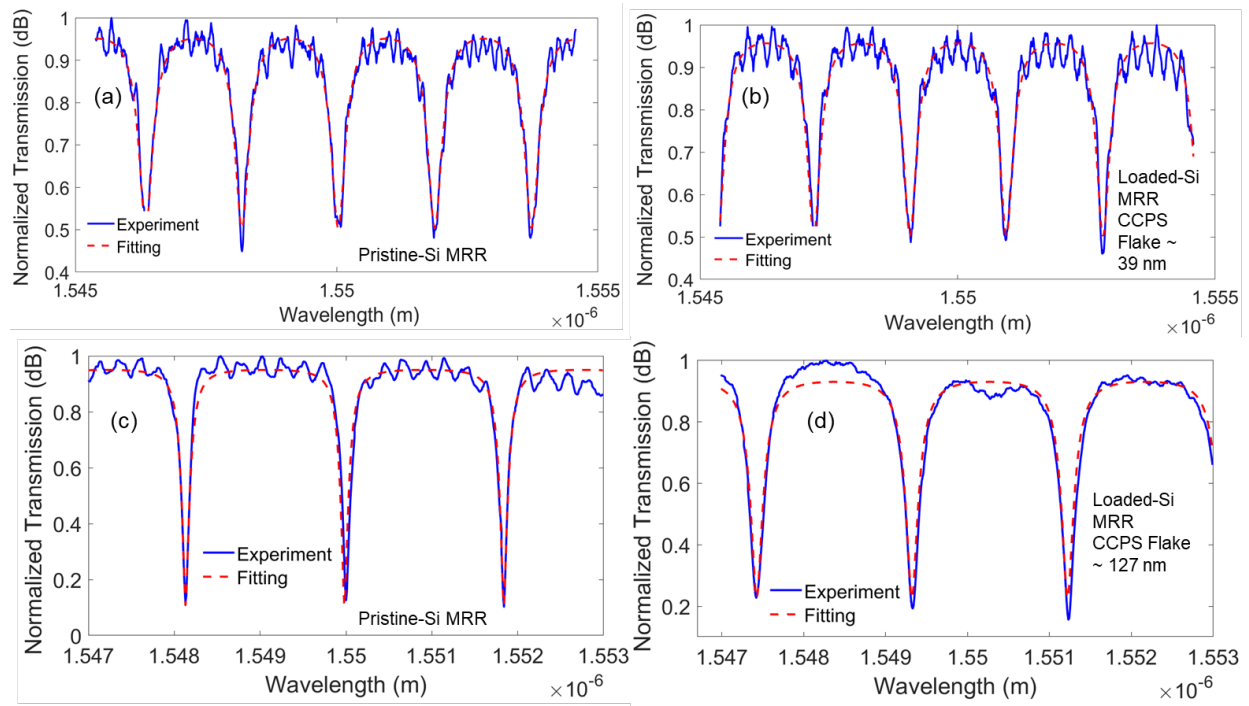

**Figure S7. Passive transmission spectra of the mirroring resonator around 1550 nm for the TE mode: (a)** Reference spectrum of the bare ring. **(b)** Spectrum after loading the ring in (a) with a ~39 nm thick CCPS flake and an interaction length of 31 μm. **(c)** Reference spectrum of another bare ring. **(d)** Spectrum after loading the ring in (c) with a ~127 nm thick CCPS flake and an interaction length of 58 μm.

**Supplement note 2**

**Electromagnet and experimental setup**

The experimental setup for magneto-optic testing utilized a low-power DC-operated cylindrical electromagnet with a moderate lifting capacity. The electromagnet is capable of exerting a maximum initial force of 50N, stabilizing at 37N. This cylindrical electromagnet features a central core and a wiring harness. When electrical current flows through the coil of wire wrapped around the metallic core, it generates a magnetic field.

The chip was placed on the flat face of the electromagnet, ensuring that the surface of the chip was parallel to the face of the electromagnet for maximum exposure to the magnetic field. The magnetic field lines exit the north pole of the electromagnet and enter the south pole, with the strongest field being perpendicular to the surface of the disk where the sample is placed. The field lines emerge from the flat face, spread out in a loop, and return through the sides and back of the electromagnet to complete the loop.

The linearity of the electromagnet was measured using a Gauss meter, which directly recorded the magnetic field values, as shown in Fig. S8(a). In our configuration, where the photonic chip was placed directly on the flat face of the electromagnet, the primary component of the magnetic field was perpendicular to the surface of the electromagnet and the plane of the chip. Given that the chip size was smaller than the diameter of the cylindrical electromagnet, any non-uniformities in the magnetic field were negligible.

This setup is illustrated in Fig. S8(b) and (c). Figure S8(b) shows the experimental arrangement with the photonic chip side-coupled and positioned on the electromagnet stage. A close-up image of the photonic chip on the electromagnet stage is presented in Fig. S8(c), with the inset showing the magnetic field profile recorded using a magnetic field viewing film based on magnetic nanoparticles.

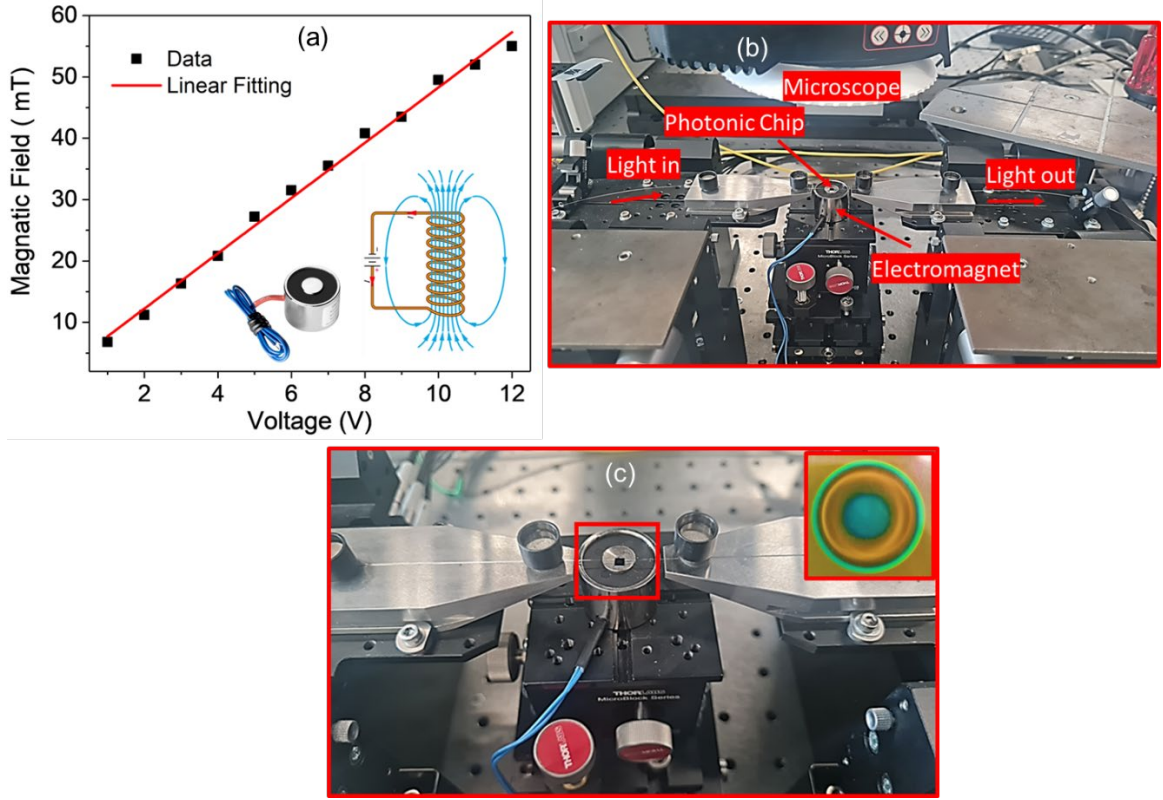

**Figure S8. Experimental setup for magneto-optic testing: (a)** Electromagnetic linearity as a function of DC applied voltage, with magnetic field values recorded using a Gauss meter. **(b)** Experimental setup featuring a photonic chip with side coupling, positioned on the electromagnet stage. **(c)** Close-up image of the photonic chip on the electromagnet stage, with the inset showing the magnetic field profile recorded using a magnetic field viewing film based on magnetic nanoparticles.

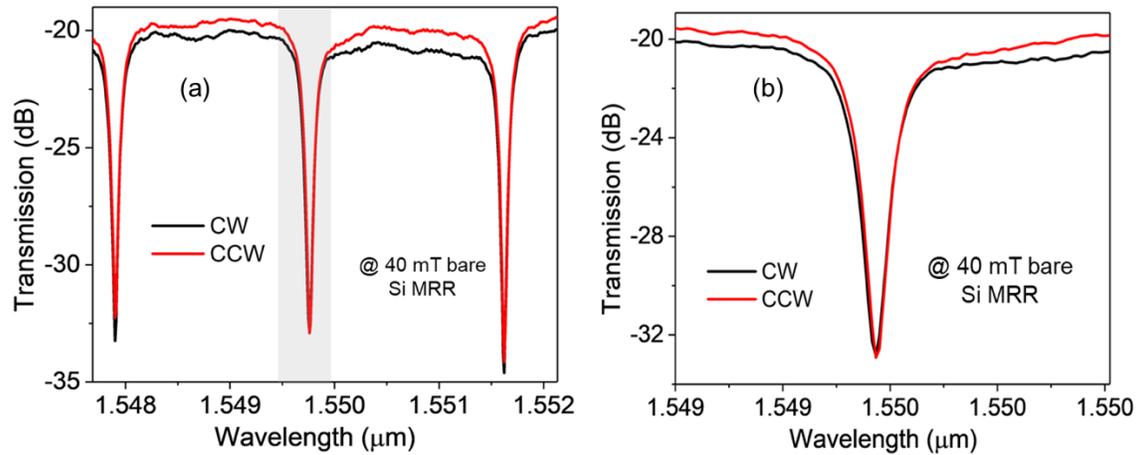

**Figure S9. Transmission spectra of the CW and CCW light propagation in a bare Si MRR under a 40 mT applied magnetic field. (a)** Wide view spectra showing three resonance peaks. **(b)** Close-up view of the peak around 1550 nm, indicating no resonance wavelength splitting (RWS) in silicon, attributed to its diamagnetic nature.

**Supplement note 3**

**Magnetic Properties and Magnetoelectric Coupling in CuCrP$_2$S$_6$ (CCPS)**

CCPS is a type-II multiferroic material that exhibits significant polarization-magnetization coupling, positioning it as a unique material for studying 2D magnetism-ferroelectricity interactions[2–4,9]. Experimental evidence has confirmed the presence of interlayer antiferromagnetism (AFM) and intralayer stripe antiferroelectricity (AFE) in its bulk form. The magnetic properties of CCPS originate from the Cr atoms, which have an incomplete 3d shell, resulting in three unpaired d electrons per atom and a corresponding magnetic moment of 3$\mu_B$ per formula unit. Below the Néel temperature ($T_N$) of around 32 K, the Cr$^{3+}$ magnetic moments align ferromagnetically within monolayers and are antiferromagnetically coupled between layers perpendicular to the van der Waals plane (See Fig.S10a). The ferroelectricity in CCPS is due to the Cu$^+$ ions, which are randomly positioned within the ferromagnetic CrS$_6$-P$_2$S$_6$ cages. At 145 K, CCPS transitions to an antiferroelectric state, with Cu ions forming a striped AFE arrangement driven by a double-well pseudopotential.

Recent studies have provided further experimental and theoretical evidence of magnetoelectric coupling in CCPS, highlighting the role of spin-orbit coupling in linking electric dipoles and spins[2,3,9,10]. This coupling indicates the existence of magnetic field-induced electric polarization and electric field-controlled magnetic orderings, even in a monolayer. Although the magnetic easy axis and spin-flop transitions in CCPS are well documented[2,4], these phenomena require further investigation for a more detailed understanding. Above the Néel temperature, CCPS is paramagnetic, with disordered spins of Cr ions. Upon cooling below $T_N$, it transitions to an AFM phase with alternating interlayer spin orientations.

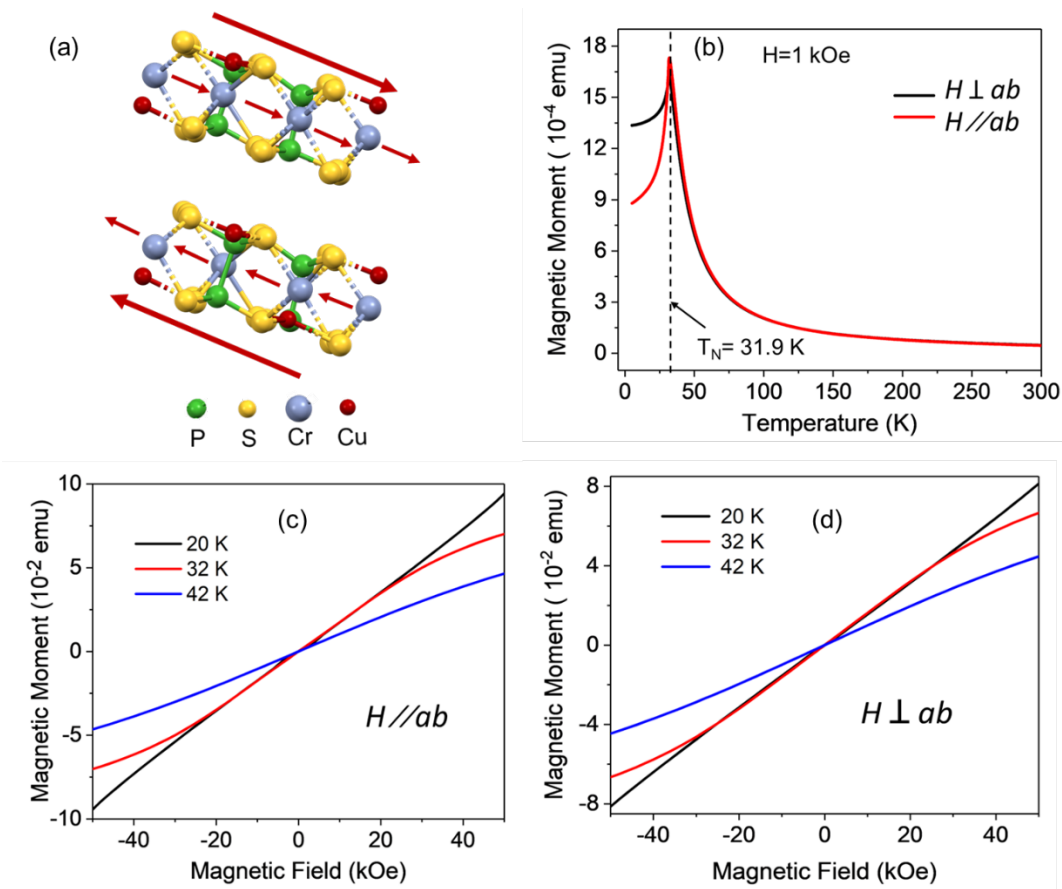

**Figure S10. Magnetic properties of CCPS: (a)** Schematic diagram illustrating the alignment of $Cr^{3+}$ magnetic moments below the Néel temperature, showing ferromagnetic alignment within monolayers and antiferromagnetic coupling between layers perpendicular to the van der Waals plane. **(b)** Temperature-dependent magnetization (M-T) curves of bulk CCPS. **(c-d)** Field-dependent magnetization (M-H) curves at various temperatures with the magnetic field applied (c) parallel to the ab crystal plane and (d) perpendicular to the ab crystal plane.

**Magnetic Measurements**

The magnetic measurements were performed using a Physical Property Measurement System (PPMS) - Quantum Design DynaCool 9T. The CCPS crystal was fixed on the sample holder and oriented parallel to different crystallographic axes at various temperatures and magnetic field settings.

The temperature-dependent magnetization (*M-T*) of CCPS is shown in Fig. S10(b) for a magnetic field of 1 kOe applied parallel and perpendicular to the *ab* planes. As the temperature decreases from 300 K, the magnetization increases monotonically, peaking around 31.9 K, indicating a transition into an antiferromagnetically ordered state with a Néel temperature ($T_N$) of 31.9 ± 1 K. A distinct cusp at $T_N$ = 31.9 K is observed in both parallel and perpendicular directions, with a sharp drop in magnetization at low temperatures along the parallel direction, suggesting that the spin order favors the *ab* plane as the easy axis of magnetization. These results align well with previous studies, which reported $T_N$ values of 31.5 K and 30.6 K[4,9], underscoring the dominant ferromagnetic intralayer coupling competing with the weaker antiferromagnetic interlayer coupling in this compound.

The field-dependent magnetization (*M-H*) at 20 K, 32 K, and 42 K with magnetic fields applied parallel and perpendicular to the *ab* plane is shown in Fig. S10(c) and S10(d), respectively. The M-H curves in both orientations exhibit nearly linear behavior, with minimal temperature dependence. This linearity and the weak temperature dependence suggest a robust ferromagnetic intralayer coupling and a weaker antiferromagnetic interlayer interaction. The ability of the spins in $CuCrP_2S_6$ to be easily polarized by an applied magnetic field supports the scenario of strong ferromagnetic intralayer coupling combined with weak antiferromagnetic interlayer coupling, consistent with previous findings showing positive Curie-Weiss values[4,9].

We conducted magnetic force microscopy (MFM) on CCPS nanoflakes of varying thickness using the NX10 Park system, which can generate an in-plane magnetic field with a maximum strength of approximately 335 Gauss (see Fig. S11). Thin flakes, specifically 20 nm and 13 nm in thickness, were selected and measured under various applied magnetic fields. The MFM results indicated that the MFM phase showed insignificant changes with the applied magnetic field. This lack of response in the MFM phase at room temperature suggests that CCPS does not exhibit ferromagnetic behavior under these conditions, indicating that the material is more likely to be paramagnetic at room temperature. However, an applied field of 335 Gauss (33.5 mT) from the bottom chuck may be insufficient to induce detectable changes in the signal.

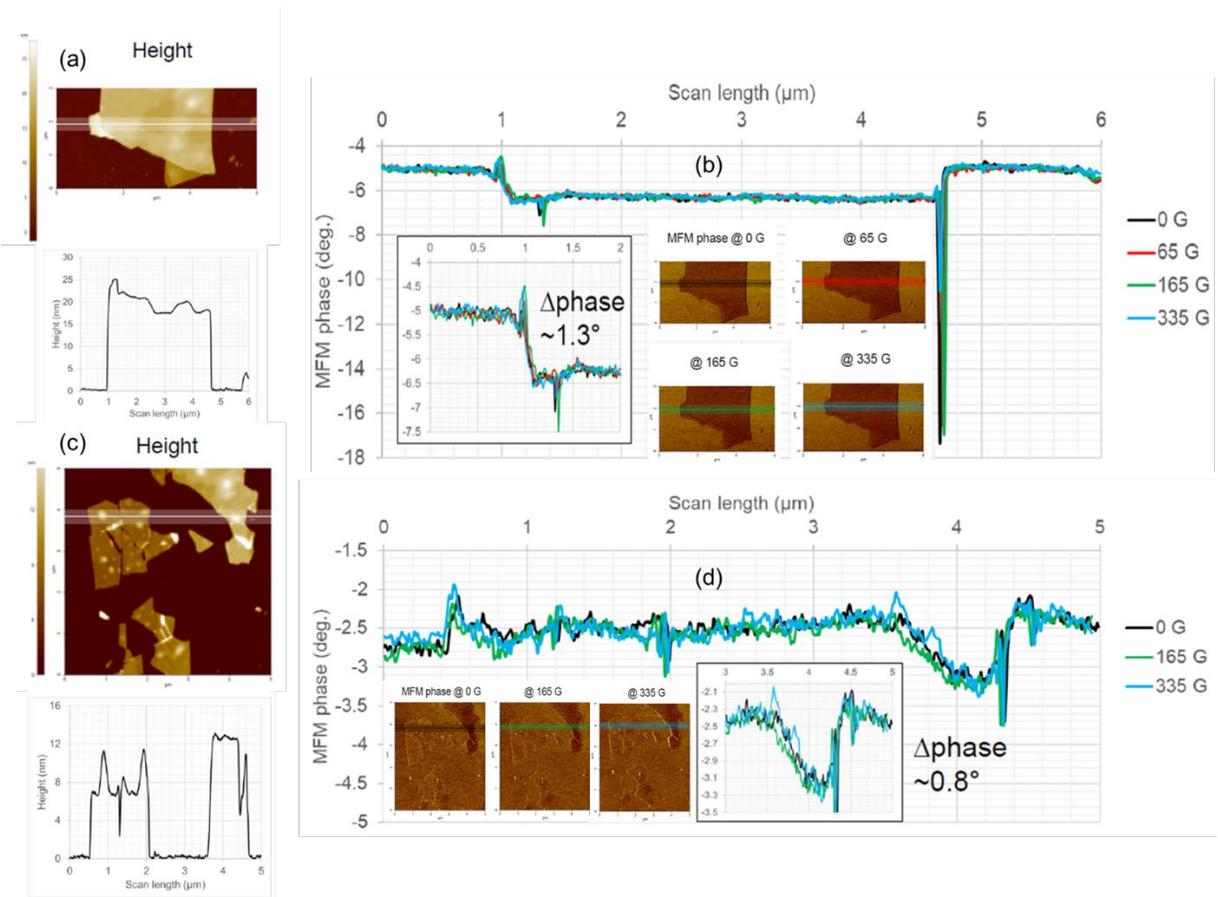

**Figure S11 Magnetic force microscopy (MFM) analysis of CCPS nanoflakes: (a)** Height sensor image displaying a nanoflake with a thickness of approximately 20 nm. **(b)** MFM phase image along lines crossing the flake, with colors indicating the applied magnetic field value. **(c)** Height sensor image displaying a nanoflake with a thickness of approximately 13 nm. **(d)** MFM phase image along lines crossing the flake, with colors indicating the applied magnetic field value.

**Supplement note 4**

**Comparative analysis, material integration and device testing**

The devices were fabricated using standard silicon-on-insulator (SOI) techniques, each chip containing 15 ring resonators. As illustrated in Fig S12(a), a pristine chip is depicted before the transfer process. Figure S12(b) provides a close-up SEM image highlighting the design details, including the 100 nm gap between the bus and the waveguide. To incorporate 2D materials into the photonic chips, we developed a unique in-house transfer method, described in our previous work[11–14].

The process began with the mechanical exfoliation of CCPS flakes using Nitto SPV224 PVC tape, followed by their transfer to a PDMS film. A gold-plated needle-based micro stamper, accurately positioned with a micro-positioner, was used to transfer the material from the PDMS. The transparency of the PDMS allowed for alignment under an optical microscope, ensuring precise flake placement on the photonic chip. Suitable flakes were identified by scanning the PDMS under a microscope, ensuring proper geometry and avoiding cross-contamination. The stamper, having a contact area larger than the flake, facilitated the precise positioning of the flake onto the target device.

This approach enabled the transfer of multiple devices. It achieves uniform CCPS layers on the rings, identifiable by consistent color contrast of the flakes, with interaction lengths ranging from 22 to 55.5 μm.

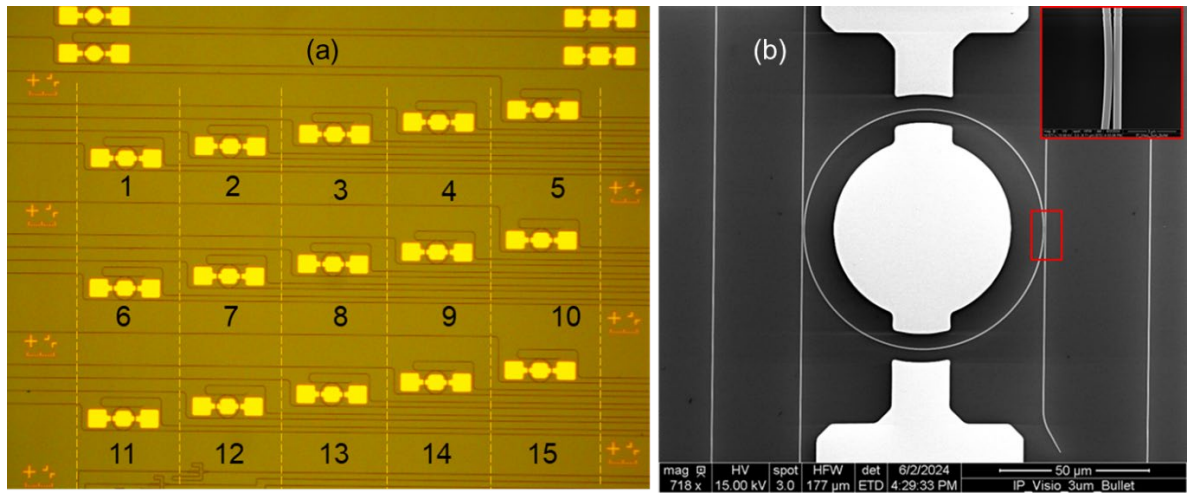

**Figure S12. Photonic chip design and fabrication: (a)** Optical microscopy image of the fabricated ring resonators prior to the transfer of CCPS, featuring 15 ring resonator devices. **(b)** SEM image providing a close-up of the fabricated resonators, with an inset highlighting the gap at the coupling region, as indicated by the red box in the resonator.

We conducted a comparative analysis to elucidate the variation in resonance wavelength splitting (RWS) and the change in effective refractive index ($\Delta n_{eff}$) across devices, as influenced by the thickness and coverage length of the flakes. Figure S13(a) presents the RWS measured around 1550 nm at 40 mT and the corresponding $\Delta n_{eff}$ values, calculated using Eq. 4 from the main paper, with $n_g$ value of 4.2 RIU extracted from simulations. This analysis highlights the impact of flake coverage and thickness on the RWS and refractive index modulation of the devices under an applied magnetic field. Measurements from at least ten devices, with varying thickness and interaction lengths as listed in Fig. S13(b), indicate that while flake thicknesses ranging from 39 to 62 nm do not significantly affect the RWS, the interaction length of the flakes does. Figure S13(c) also shows SEM images of the various transferred devices utilized in our study.

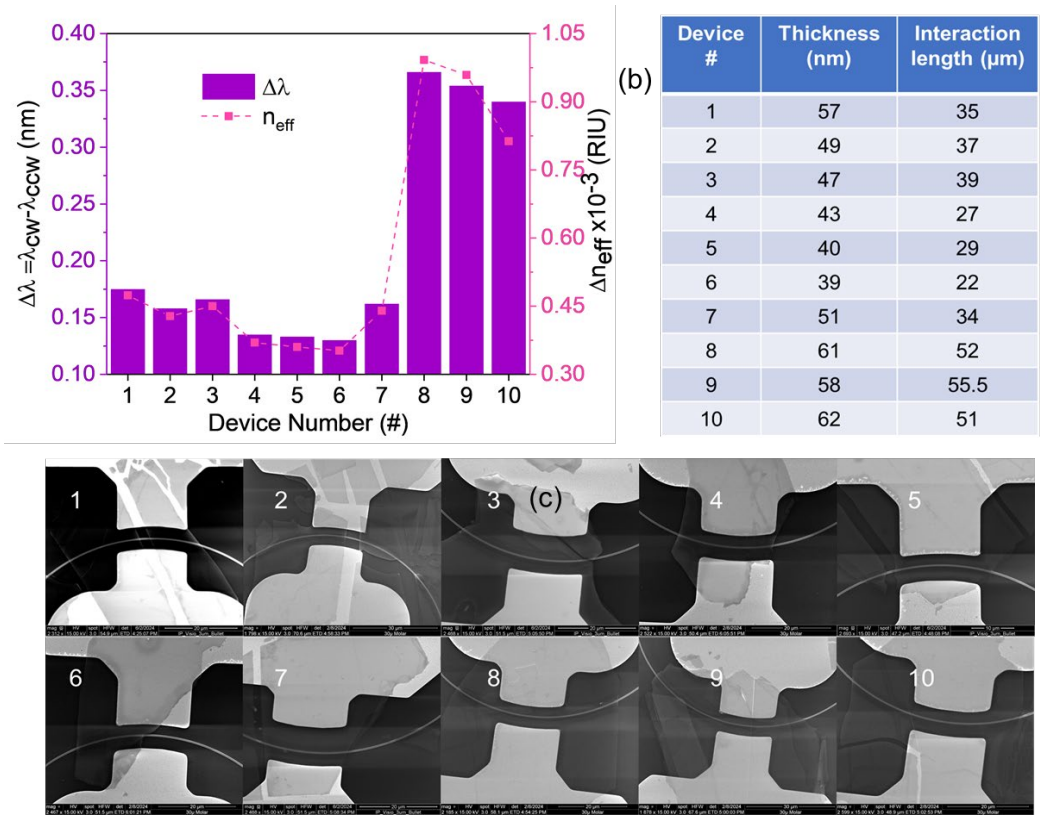

**Figure S13. Comparative analysis: (a)** Resonance wavelength splitting (RWS) between CW and CCW light propagation and change in effective refractive index ($\Delta n_{eff}$) for 10 tested devices with thicknesses ranging from 22 nm to 62 nm and interaction lengths from 22 to 55.5 µm. **(b)** Thickness and interaction lengths of the tested devices. **(c)** Scanning electron microscopy (SEM) images of the transferred and tested devices during our experiments.

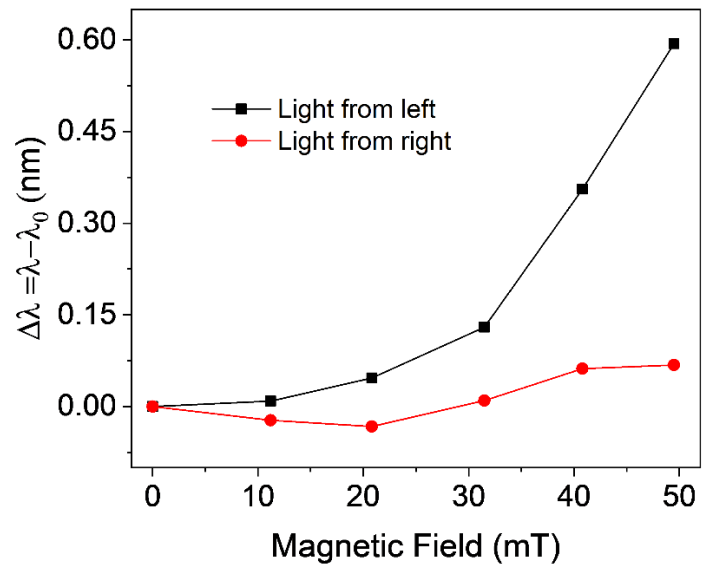

**Figure S14. Resonance shift as a function of the magnetic field.** The resonance peak position at zero magnetic field, $\lambda_0$, serves as a reference. The light was launched in either the right or left direction, and the resulting resonance shift was recorded as the magnetic field increased. Notably, the shifts induced when the light was launched from the left differed in magnitude from those observed when launched from the right, indicating a clear deviation from the symmetric shifts expected under thermal effects.

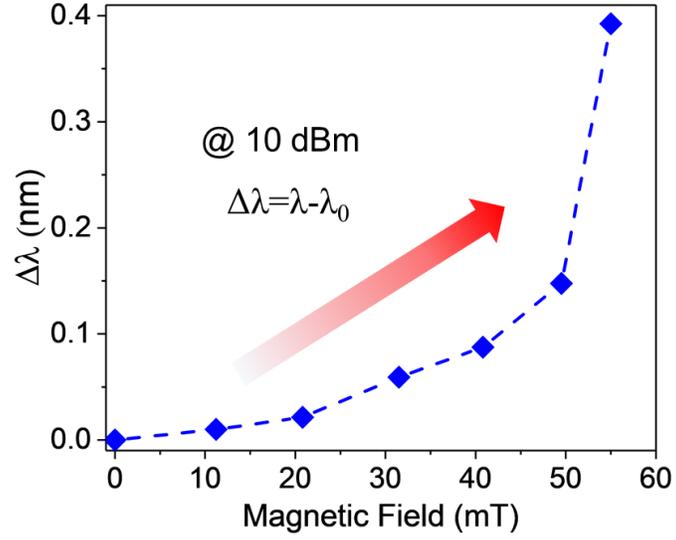

**Figure S15. Magneto-optic characteristics of the hybrid CCPS/Si micro-ring resonator (MRR). (a)** Resonance peak positions at a constant input laser power of 10 dBm, launched from the left, as the magnetic field varies from 0 mT to 55 mT. Here, $\lambda_0$ denotes the resonance peak position at 0 mT, and $\lambda$ represents the resonance peak position at different magnetic field strengths.

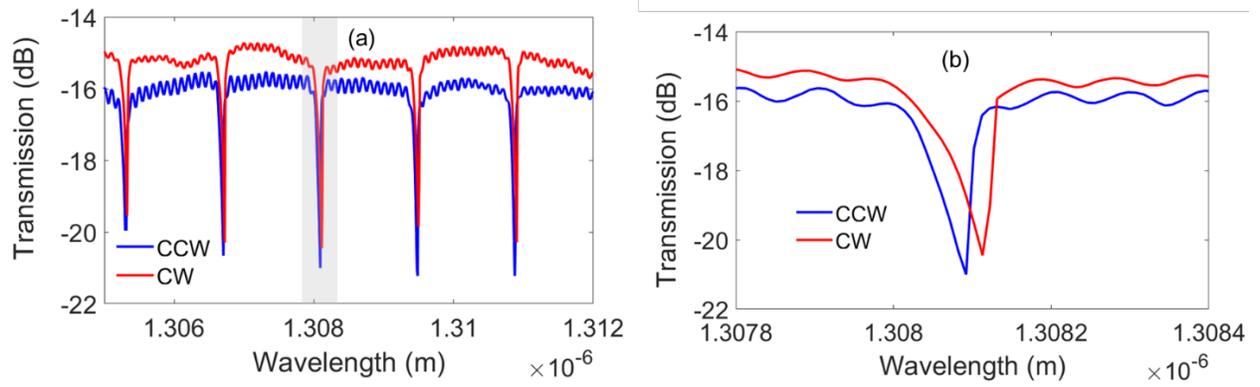

**Figure S16. Transmission spectra of clockwise (CW) and counterclockwise (CCW) light propagation for a 1310 nm centered laser under a 40 mT magnetic field. (a)** Wide scan view. **(b)** Close-up view highlighting the nonreciprocal response around 1310 nm.